# Nonpolar *p*-GaN/*n*-Si heterojunction diode characteristics: A comparison between ensemble and single nanowire devices

Avinash Patsha, Ramanathaswamy Pandian, Sandip Dhara and A. K. Tyagi

*Surface and Nanoscience Division, Indira Gandhi Center for Atomic Research, Kalpakkam-603102, India*

Email: avinash.phy@gmail.com, rpandian@igcar.gov.in

**Abstract**

The electrical and photodiode characteristics of ensemble and single *p*-GaN nanowire and *n*-Si heterojunction devices were studied. Ideality factor of the single nanowire *p*-GaN/*n*-Si device was found to be about three times lower compared to that of the ensemble nanowire device. Apart from the deep-level traps in *p*-GaN nanowires, defect states due to inhomogeneity in Mg dopants in the ensemble nanowire device are attributed to the origin of high ideality factor. Photovoltaic mode of ensemble nanowire device showed an improvement in the fill-factors up to 60 % over the single nanowire device with fill-factors up to 30 %. Reponsivity of the single nanowire device in photoconducting mode was found to be enhanced by five orders, at 470 nm. The enhanced photoresponse of the single nanowire device also confirms the photoconduction due to defect states in *p*-GaN nanowires.

# 1. Introduction

GaN-based nanowire (NW) electronic and optoelectronic devices such as high electron mobility transistors, LEDs, lasers, photodetectors and solar cells showed remarkably superior performances with respect to their thin film counterparts [1-4]. Combining with other III-nitrides, GaN-based single and ensemble NW photoconductors and photodiodes paved the way for realizing photoresponse for wide wavelength range [5,6]. Besides, the absence of polarization-related electric fields in nonpolar III-nitride-based nanostructures provides a wider range of absorption wavelengths along with higher absorption probability compared to polar components [7,8]. UV photodetectors with ultrahigh photocurrent gains up to $10^5$ are demonstrated with nonpolar GaN and GaN/AlN single NWs [9-11]. Further improvements on the functionalities of the NW devices could be achieved by forming heterojunctions with other electronic and optoelectronic materials. Earlier studies showed the feasibility of integrating GaN NW devices with other semiconducting materials such as Si and ZnO [12-17]. There is also report on the optoelectronic property of heterojunction devices (GaN/ZnO) that can be controlled by the interface polarity [13]. The recent report on integration of III-V nanostructures on Si show the possibility of vertical electronic and optoelectronic nano devices [18]. However, electrical and optical characteristics of such heterojunction devices can be influenced by several device processing parameters which enhance the integration complexity.

Generally GaN based *p-n* junctions are found to have high diode ideality factors deviating from normal range of 1-2 according to Sah–Noyce–Shockley model [19,20]. Such high values are attributed to various origins including quantum barriers in GaInN/GaN LEDs [20], sum of ideality factors of individual rectifying junctions in *p-n* junctions of GaN and AlGaN/GaN superlattice [21], and band gap states formed by crystalline imperfections [22]. A comparative study made between *p-n* diodes composed of GaN thin film and nanostructures revealed that the ideality factor could reach values as high as 20 [23]. Since it is possible that heterojunctions of GaN NWs can influence the carrier transport mechanism, ideality factors of the devices are therefore expected to vary. However, there is hardly any report on the variations in electrical and optical characteristics of such heterojunction devices.

In this article, we present a comparative study on ensemble and single nonpolar Mg-doped *p*-GaN NW and *n*-Si heterojunctions. Particularly, photovoltaic- and photoconducting-mode characteristics of uniform-sized *p*-GaN NWs grown on *n*-Si(111) by chemical vapor deposition (CVD) method are analyzed and the key parameters including ideality factors, fill factors and responsivity of ensemble and single NW heterojunction devices are estimated for comparison.

## 2. Experimental details

### 2.1 Nanowire synthesis

Mg-doped nonpolar GaN NWs were synthesized by atmospheric pressure chemical vapor deposition (APCVD) technique via the catalytic vapor-liquid-solid (VLS) process. The detailed growth process can be found elsewhere [24]. In brief, Ga metal (99.999%, Alfa Aesar, as Ga source), $NH_3$ gas (99.999%, as N source), and $Mg_3N_2$ (Alfa Aesar, as Mg source) were used for synthesizing Mg-doped GaN NWs. Nanowires were grown at 900 °C for one hour by purging 10 sccm of $NH_3$ (reactant gas) and 20 sccm of Ar (carrier gas) on *n*-Si(111) substrates containing Au nanoparticles. Prior to Au catalyst deposition, the substrates were cleaned using standard RCA method to remove any organic, native oxide and metallic contaminations present on the surface of the substrates. Mg dopants, for *p*-type conduction in GaN NWs, were activated by thermal annealing of the as-grown samples in UHP $N_2$ atmosphere at 750 °C for 30 minutes. Samples of undoped GaN nanowires (*i*-GaN) on *n*-Si were also synthesized for the comparison.

*2.2 Characterizations and device fabrication*

Morphological features of the NWs were examined by field-emission scanning electron microscope (FE-SEM, SUPRA 55 Zeiss). The photoluminescence (PL) studies of *p*-GaN NWs were carried out at 80 K with an excitation wavelength of 325 nm of the He−Cd laser. The spectra were collected using 2400 lines $mm^{-1}$ grating and thermoelectrically cooled CCD detector. Ensemble *p*-type GaN NW/*n*-Si heterostructure devices were fabricated by depositing Au (100 nm)/Ni (25 nm) interdigitated electrodes (active area of 12 $mm^2$) onto the NWs using thermal evaporation and establishing an electrical contact with Si substrate by GaIn eutectic. Prior to the deposition of interdigitated electrodes, the substrate of ensemble nanowires was coated with PMMA and then the top surface was exposed to acetone to remove a certain thickness of the PMMA layer to expose the nanowire tops. After depositing Au (100 nm)/Ni (25 nm) contacts on nanowires, the PMMA was completely removed by dissolving in acetone. The devices were annealed at 400 °C for 5 min in inert atmosphere to ensure Ohmic contacts. Similarly, ensemble nanowire devices consisting of *i*-GaN on *n*-Si were also fabricated by depositing Au/Ti/Al/Ti interdigitated electrodes for the reference purpose. A single *p*-type GaN NW/*n*-Si heterostructure device was fabricated by dual-beam (focused ion beam and electron beam) technique (FIB-SEM, AURIGA, Zeiss) on $SiO_2$ (300 nm) substrate patterned with Au/Cr pads (200 $\mu m^2$). NWs were separated from the growth substrate and transferred to another $SiO_2$ (300 nm) coated Si substrate which contained pre-patterned 200 square microns Au/Cr pads. The NW was electrically connected to Au/Cr pads by Pt nano-strips deposited by FIB technique. Prior to the FIB deposition, small pads (1 X 0.6 µm) of Pt were deposited on both ends of the nanowire by electron beam in order to avoid the possible damage of the NWs by $Ga^+$ ion beam. Care was taken to minimize the spread of Pt out of the marked region, and for that a minimum possible ion current (5 pA at 30 kV) was used for deposition. A selective removal of $SiO_2$ just beneath the centre of NW was performed by ion beam to allow junction formation between *p*-GaN and *n*-Si. Agilent source measure units (B2902A and B2911A) were used to study the current-voltage (I-V) characteristics of the fabricated devices. The photodiode characteristics of ensemble and single *p*-GaN NW/*n*-Si heterostructure devices were studied in forward and reverse bias configuration under illumination of 470 and 530 nm wavelengths. The power of the illumination corresponding to different wavelength sources is normalized by considering the spot size (radius: 9.6 mm) of the illumination and the active area of both the ensemble and single nanowire heterojunction device.

### 3. Results and discussions

*3.1. Morphological and optical characteristics of p-GaN NWs*

SEM micrographs show the size distribution and arrangement of nanowires (Figure 1). Average diameter of the NWs was estimated to be 70 ($\pm$10) nm. Particles at the tips of NWs confirm the growth involving VLS process (inset in figure 1). Our earlier report on structural analysis by high resolution transmission electron microscopy revealed the predominant growth direction of [10-10] for Mg-doped GaN NWs in the wurtzite phase [24]. PL spectra of ensemble *p*-GaN NWs were collected at different locations on the NWs sample cooled to the temperature of 80 K. Detailed PL studies of different concentrations of Mg dopants in GaN NWs can be found elsewhere [24]. PL spectra at 80 K show the broad band ultraviolet luminescence (UVL) around 3.0-3.3 eV and low intensity blue luminescence (BL) peak around 2.85 eV (figure 2). Both the UVL and the BL peaks are characteristics of Mg doped GaN[25]. The presence of cubic phase as origin of the PL peaks below 3.4 eV in Mg doped GaN nanowires [26,27] ~~is~~ was ruled out as the elaborate structural studies did not show any indication for the same. [24]. The BL peak is assigned to transitions between shallow acceptors ($Mg_{Ga}$) and deep donors ($V_N$). The UVL band is due to donor-acceptor pair transitions involving shallow acceptors ($Mg_{Ga}$) and deep donors. Intensity variations in BL peak and shift in the UVL position show the inhomogeneous dopant distribution in ensemble *p*-GaN NWs of the sample.

### *3.2. Ensemble p-GaN NW/ n-Si heterojunction device*

The I-V characteristics of undoped GaN (*i*-GaN) NWs on *n*-type Si (*i*-GaN/*n*-Si) heterojunction show a typical Schottky behavior under ± 4 V bias (figure 3(a)). A simple 3D schematic view of the device (*p*-GaN/*n*-Si) showing the interdigitated electrodes on ensemble nanowires is depicted in figure S1(a) of the supplementary information. On applying ±4 V bias, the I-V characteristics of *p*-GaN/*n*-Si device showed a *p-n* diode behavior under dark (figure 3(b)). A schematic for cross-section view of the heterojunction device in forward bias configuration is shown in inset of figure 3(b). In forward bias, threshold voltage of the device is found to be ~ 0.7 V and rectification ratio under dark was found to be ~ $10^4$ at ± 4 V. The I-V characteristics of a *p-n* heterojunction can be described by the equation, $I = I_0 [\exp(\frac{qV}{nkk_BT}) -1]$, where $I_0$ is reverse current, $q$ is electron charge, $k_B$ is Boltzmann constant, $T$ is temperature and $n$ is ideality factor [28]. In low forward bias range of 0.2-0.5 V (figure 3(b)), calculated ideality factor is found to be 2. This shows that carrier transport is limited by generation-recombination processes [19]. Reverse current in this region is about $10^{-9}$ A. In voltage range of 0.5-1.0 V, a very high ideality factor value of 10 was obtained and the reverse current was about $10^{-6}$ A. In general, deviation of *n* values from 1-2 of ideal *p-n* diodes could be understood by studying the temperature dependence of *n* and behavior of metal-semiconductor contacts.

In present study, contacts for both *p*-GaN and *n*-Si were carefully formed with appropriate metals for individual semiconductors. The I-V characteristics of metal-semiconductor junctions (figures 4(a) and 4(b)) were studied separately and found to be Ohmic. Hence the contribution of metal-semiconductor junctions to *n* values is expected to be minimum. High concentrations of Mg dopants could also reduce the series resistance of *p*-GaN NWs used for heterojunction device. However, carrier diffusion through interfaces between NWs influencing the ideality factor cannot be ruled out because of the random orientations of *p*-GaN NWs. Dependence of *n* values on temperature was studied for ensemble *p*-GaN/*n*-Si heterojunction under dark condition (figure 3(c)). The I-V characteristics showed a decrease in the *n* value from 10.6 to 9.5 while increasing temperature from 300 to 425 K. Small reduction in the *n* value indicates that the measured electrical parameters are independent of contacts and further confirms the Ohmic nature of contacts.

Unusually high ideality factors in GaN *p-n* diodes are suggested to originate from band gap states formed by deep-level impurities such as hydrogen complexes with Mg or native defects like nitrogen vacancies. In III-nitride based AlGaN/GaN [21], and GaInN/GaN multiple quantum well light-emitting diodes [20], the origin was found to be GaN quantum barriers and individual rectifying junctions. In present study, ensemble *p*-GaN/*n*-Si heterojunctions were fabricated in-situ during the growth of Mg-doped *p*-GaN NWs on *n*-Si. Therefore any excess and inhomogeneous incorporation of Mg dopants, while growing GaN NWs on Si, could modify the electrical properties of heterojunction. In fact PL study at different locations of *p*-GaN NWs used in ensemble NW device showed shift in donor acceptor pair peak position (figure 2) indicating the inhomogeneity in Mg doping. Thus apart from deep-level impurities formed by Mg in *p*-GaN

NWs [22], defect states due to inhomogeneity in doping of different GaN NWs could also increase the ideality factor of ensemble NW device consisted of large junctions.

### *3.3. Photodiode characteristics of ensemble NW heterojunction device*

Ensemble nonpolar *p*-GaN NW/*n*-Si heterojunction device was examined to probe its photodiode characteristics. Photoresponse characteristics were studied under the illumination of 470 and 530 nm wavelengths with a bias voltage of ± 4 V. When the device was operated in the photovoltaic (PV) mode observed open circuit voltage ($V_{oc}$), short circuit current ($I_{sc}$) and fill factor (FF) for exposure with λ = 470 nm were 300 mV, 1 µA and 54 %, respectively. While for exposure with λ = 530 nm, $V_{oc}$, $I_{sc}$, and FF are 310 mV, 0.8 µA and 56% respectively (figure 3(b)). When the device was operated in the photoconducting (PC) mode in reverse bias, a change in reverse bias current was observed for both the wavelengths (figure 3(b)). Temporal response of the device (figure 3(d)) showed that the maximum responsivity ($R_\lambda$) was 69 mA/W at 470 nm whereas 18 mA/W at 530 nm. The corresponding external quantum efficiencies were 18.2 % and 4.2 % for 470 and 530 nm, respectively. All data corresponding to electrical and photodiode characteristics are tabulated in supplementary information (Table S1). Under the equilibrium condition, the energy bands of both the semiconductors get aligned such that the Fermi levels coincide on both sides of the heterojunction. Formation of heterojunction between *p*-GaN NWs and *n*-Si shifts Fermi level on both sides (see schematic band diagram in supplementary figure S1(b)). PV effect at zero bias showed separation of photogenerated carriers due to built-in potential at the junction between *p*-GaN NWs ($E_g$ = 3.47 eV) and *n*-Si ($E_g$ = 1.12 eV) [16]. Carrier generation with photon energies less than the band gap of GaN (3.47 eV ≈ 357 nm) indicates the possible injection of charge carriers from *n*-Si. Whereas, when the device was operated in the PC mode at a reverse voltage of 4V, the enhanced photoresponse for 470 nm, compared to 530 nm (figure 3(d)), showed the possibility of larger photocurrent generation due to defects formed by Mg doping in *p*-GaN NWs. Earlier studies on spectral response of PC measurements on GaN NWs have revealed similar defect related photocurrent for sub-band gap excitations [29]. In fact, PL studies on *p*-GaN NWs in present report, showed the available electronic states which might be responsible for photocurrent (figure S2 in supplementary information). With increased reverse bias voltage in the PC mode, the enhanced collection of photo generated carriers in *p*-GaN NWs due to defects could enhance the responsivity.

### *3.4. Single p-GaN NW/ n-Si heterojunction device*

Electrical and photodiode characteristics of ensemble *p*-GaN/*n*-Si heterojunction device were compared with single *p*-GaN NW/*n*-Si heterojunction device (figure 5). Magnified view of a 70 nm diameter *p*-GaN NW with Pt electrodes is shown in inset of figure 5. A cross-sectional schematic view of the device connected in a forward bias is described in the supplementary information (figure S3). With an applied bias of ±1 V between (one of the) Au/Cr contact pads of *p*-GaN NW and top-contact of *n*-Si, the I-V characteristics of single NW device showed a *p-n* diode behavior under dark (figure 6(a)). Forward bias current was limited (with a maximum value) to avoid any accidental breakdown of the device due to joule heating and/or spurious electrical spikes. In forward bias, turn-on voltage of the device was found to be ~0.3 V and rectification ratio under dark was higher than $10^4$ at ±1 V.

Using the diode equation, ideality factors were calculated for the single NW device with forward bias between 0.1 and 0.5 V under dark. In low bias range, 0.1-0.2 V, the ideality factor was 2.8 and reverse bias current was $2 \times 10^{-10}$ A (figure 6(a)). The magnitude of ideality factor indicated the deviation of carrier transport from recombination processes. In high bias range, 0.2-1 V, the ideality factor was ~ 3.7 and reverse current was ~ $3 \times 10^{-10}$ A. Ideality factor of single

NW device is significantly lower than the ensemble NW device in high bias range. The Ohmic contact behavior of Pt/*p*-GaN NW was studied under dark (figure 4(c)). FIB-deposited Pt contact is reported to be Ohmic for GaN NW with diameter below 100 nm [30]. Temperature dependence of the ideality factor of single NW device was also studied under dark (figure 6(b)). The calculated *n* values show a decreasing trend from 3.5 to 2.6 with an increase in temperature from 300 to 425 K. A little decrease in *n* value could be due to slight improvement in Pt/*p*-GaN contact while increasing the temperature. In single NW device, role of defect sates due to the inhomogeneity in dopant could be negligible. However, deviation of *n* value (>2) could be due to deep-level impurities formed by Mg in *p*-GaN NW as mentioned earlier. Reduction in the magnitude of ideality factor in single NW heterojunction compared to ensemble NW device could be due to the reduction in number of NWs with defect states caused by inhomogeneously doped *p*-GaN NW.

### 3.5. Photodiode characteristics of single NW heterojunction device

Single *p*-GaN NW/*n*-Si heterojunction device (active area of $\sim 7 \times 10^4$ nm$^2$) was analyzed to explore its photodiode characteristics. Similar to ensemble NW device, the photoresponse was tested under the illumination of 470 and 530 nm wavelengths. When the device was operated in the PV mode (figure 6(a)), maximum $V_{oc}$ of 100 mV was observed for 470 nm while for 530 nm, $V_{oc}$ was 70 mV. $I_{sc}$ under the exposure of 470 and 530 nm were found to be 3.5 and 1.1 nA, respectively. Fill factors of 32 and 31 % were determined for 470 and 530 nm, respectively. These values are comparable to the reported values of ~38% for ensemble *p*-GaN NW/*n*-Si devices [16]. When the device was operated in the PC mode, a change in reverse bias current was observed for both wavelengths (figure 6(a)). Temporal response of the device shows (figure 6(c)) the maximum responsivity of 1675 A/W at 470 nm and 197 A/W at 530 nm. These high responsivity values yield high photocurrent gains. All the relevant data for single NW device are tabulated in Table S1 of the supplementary to compare with the ensemble NW device. High responsivity and photocurrent gain values for above band gap excitation energies are reported earlier for single GaN NW photoconductors [9-11] as well as for photodiodes [12]. In present study, however the excitation energies were less than the band gap energy of GaN. As mentioned in case of ensemble NW device, high photoresponse for below sub-band gap excitation energies could be due to the carriers generated by Mg-doping-induced defect states in *p*-GaN NWs [29]. Besides, the geometry of single NW device could also allow *p*-GaN NW to be fully exposed to incident photons and contribute for surface conduction, therefore enhancing the photoresponse in the PC mode [12]. A drastic increase in responsivity of the single NW device compared to ensemble NW device confirms the role of defect states in the carrier conduction mechanism. However, the role of effective absorption cross-section being larger than the physical nanowire cross-section may not be completely ruled out for the observed high responsivity of a single nanowire device.

## 4. Conclusions

In conclusion, ensemble and single Mg-doped nonpolar *p*-GaN nanowire/*n*-Si heterojunction devices were fabricated and their electrical and photodiode characteristics were compared. High ideality factor of 10 was observed for an ensemble *p*-GaN NW/*n*-Si device. Defect states due to deep-levels and inhomogeneity in Mg dopants are attributed to the increase in ideality factor. Photovoltaic mode of ensemble NW device showed an improvement in the fill-factors up to 60 % over the single NW device with fill-factors up to 30 %. Responsivity of the single NW device in photoconducting mode was found to be enhanced by five orders. The enhanced photoresponse of the single NW device also confirms the photoconduction due to defect states in *p*-GaN nanowires.

**Acknowledgments**

One of the authors, Avinash Patsha, acknowledges Department of Atomic Energy for the financial support.

**Figures:**

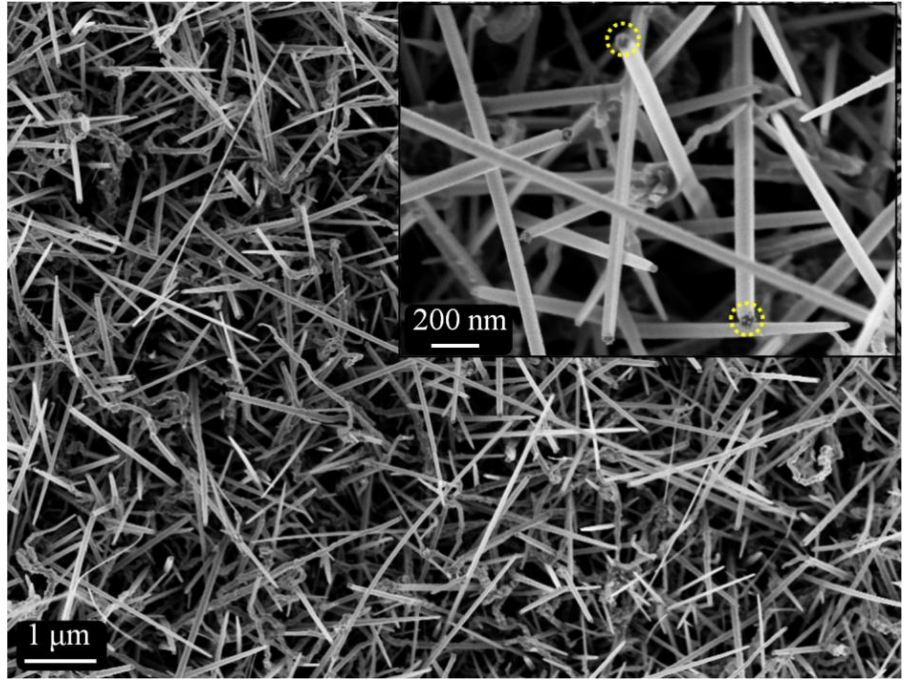

Figure 1. FE-SEM images of as-grown Mg-doped *p*-GaN nanowires on n-Si(111) substrate. Inset shows Au catalyst particles (encircled) at the tip of the wire

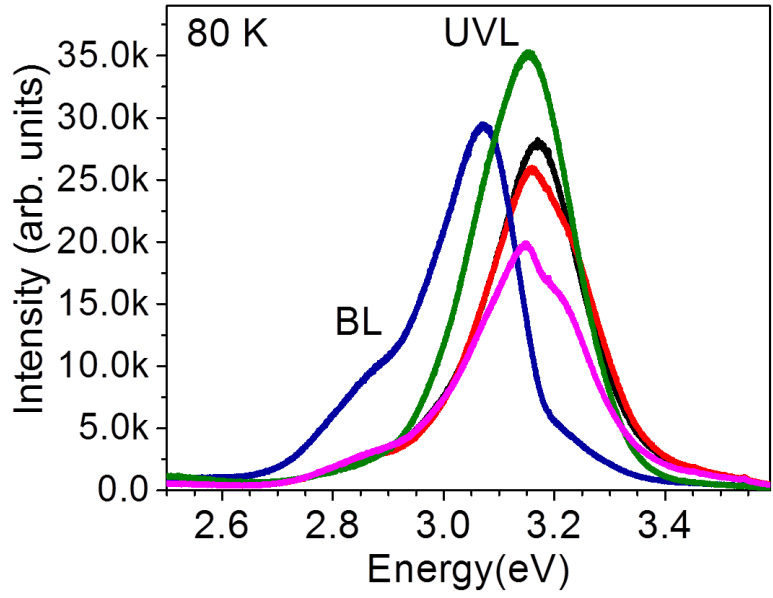

Figure 2. Photoluminescence spectra of ensemble *p*-GaN nanowires recorded at different locations on the sample.

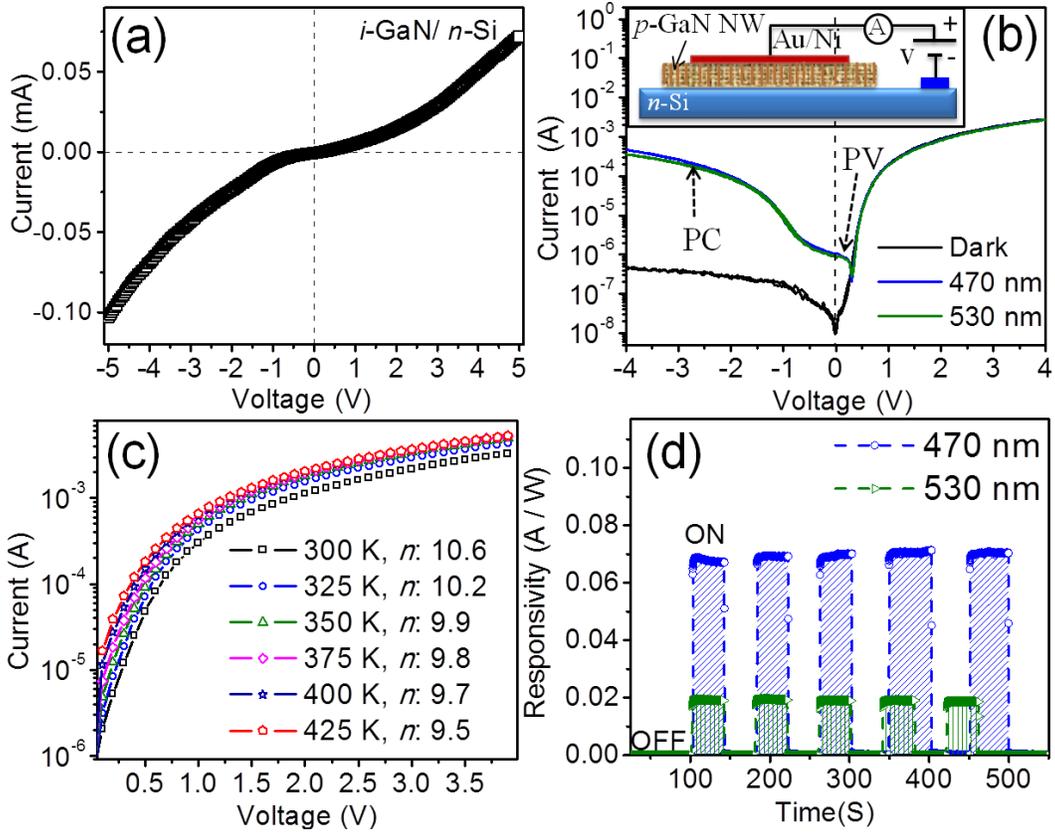

Figure 3. (a) Typical I-V characteristics undoped GaN NWs on *n*-type Si (*i*-GaN/*n*-Si) heterojunction device under dark; (b) I-V characteristics of ensemble NW of *p*-GaN/*n*-Si heterojunction device under dark and illumination of different wavelengths. Inset shows the schematic of the heterojunction device in forward bias; (c) Temperature dependent I-V characteristics of the ensemble NW device under dark; (d) Temporal photoresponse of the ensemble NW device under the illumination of 470 and 530 nm with light ON and OFF

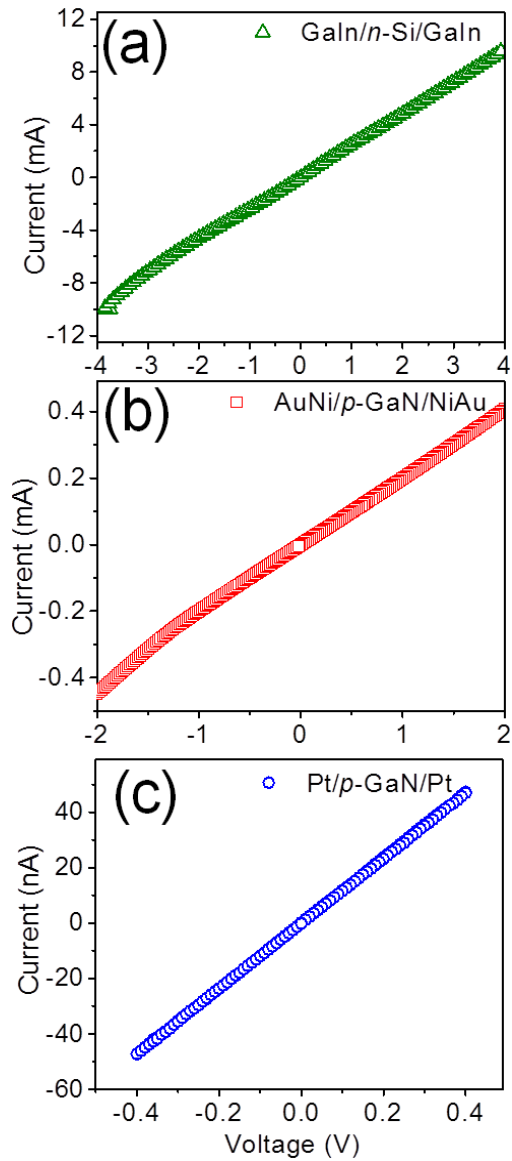

Figure 4. I-V characteristics of (a) GaIn/*n*-Si/GaIn, (b) AuNi/*p*-GaN/NiAu and (c) Pt/*p*-GaN/Pt under dark at 300 K showing Ohmic nature of the contacts.

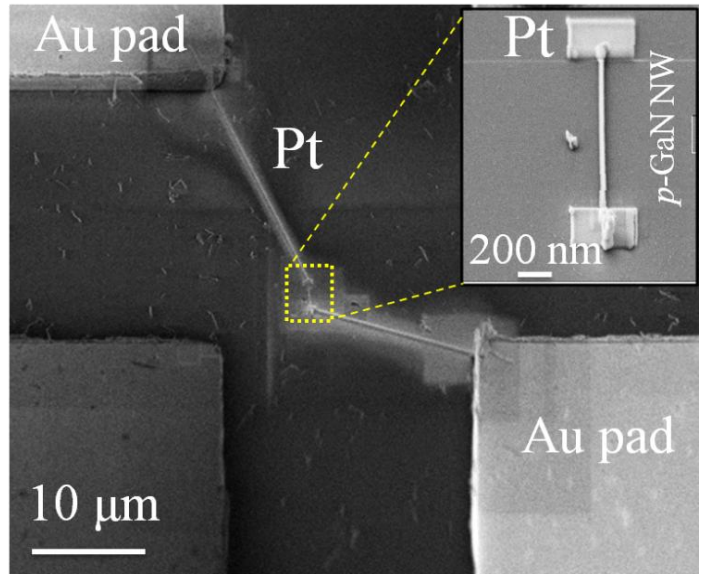

Figure 5. FE-SEM micrograph of single *p*-GaN NW/*n*-Si heterojunction device. Inset shows the magnified view of *p*-GaN NW with electron beam deposited Pt (primary layer) electrodes

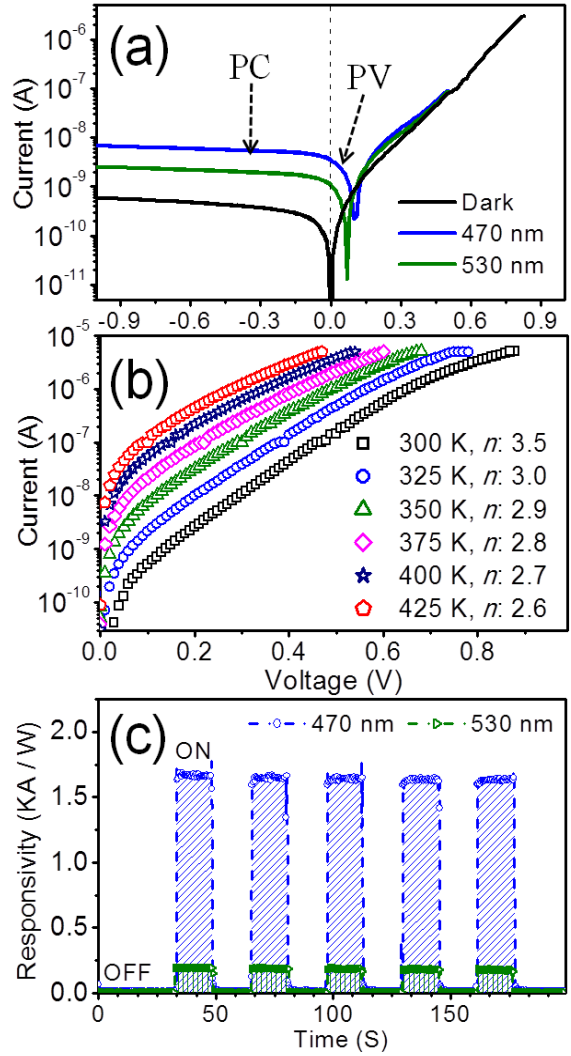

Figure 6. (a) I-V characteristics of single *p*-GaN NW/*n*-Si heterojunction device under dark and illumination of different wavelengths; (b) Temperature dependent I-V characteristics of the device under dark; (c) Temporal photoresponse of the device under illumination of 470 and 530 nm wavelengths

Supplementary Information:

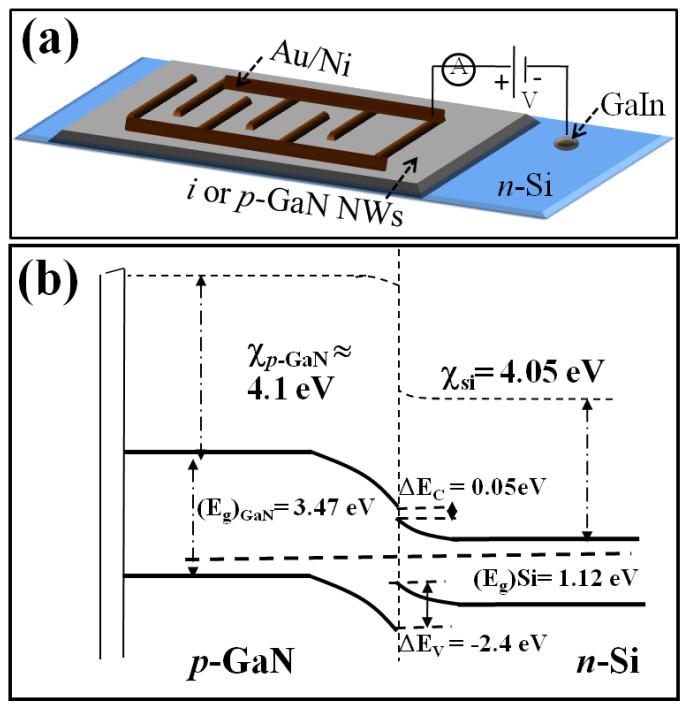

Figure S1. (a) 3D schematic view of the inerdigitated electrodes of ensemble nanowire (NW) device in (b) The formation of heterojunction between *p*-GaN NWs and *n*-Si with Ohmic contacts on both sides.

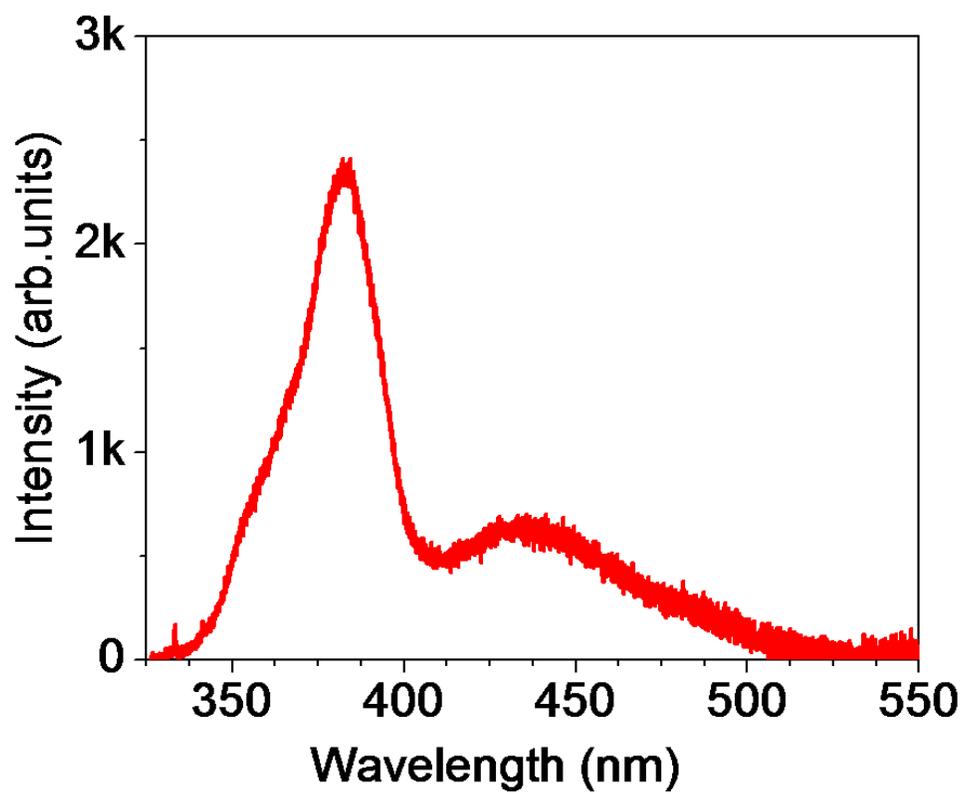

Figure S2. Photoluminescence spectra of *p*-GaN nanowires showing possible electronic states available around 470 nm.

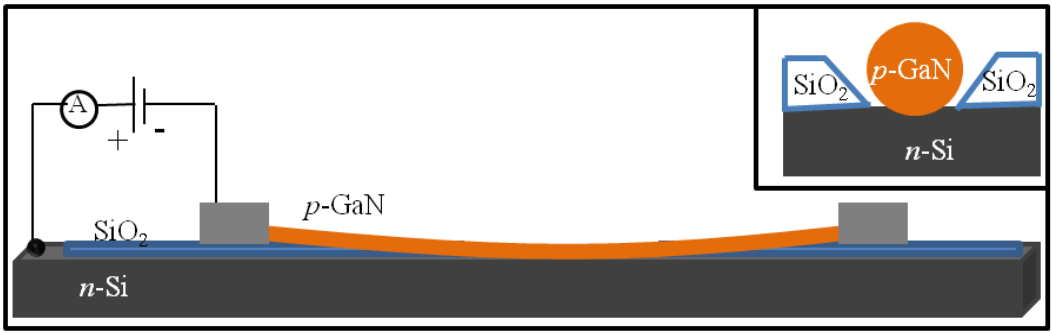

Figure S3. A schematic cross-sectional side-view (along the nanowire) of a single nanowire *p*-GaN/*n*-Si heterojunction device in the forward bias configuration. Inset shows the cross-sectional front-view of the device.

The etching of SiO$_2$ layer beneath the nanowire was carried out by tilting the nanowire substrate with respect to the ion beam and focusing at substrate surface near one side of the nanowire. This allowed the SiO$_2$ to be etched in a wedge shaped trench parallel to the wire axis. Similar procedure was repeated on the other side of the nanowire. A short exposure of ion beam (with low beam current; 5 pA) on nanowires above the wedge shaped trench bends the wire in to the bow shape until it touches the bottom of wedge shaped trench. The working conditions of the FIB are, working distance; 5.1 mm, angle between ion beam and substrate; 54°, Accelerating voltage; 30 kV.

TABLE S1. Comparison of electrical and photodiode characteristics of ensemble and single *p*-GaN NW/*n*-Si heterojunctions.

| Measured parameters | Ensemble NWs device | Single NW device |
|---|---|---|
| Electrical characteristics: | | |
|     Threshold voltage ($V_{th}$) | 0.7 V | 0.3 V |
|     Rectification ratio | $10^4$ | $10^4$ |
|     Ideality factors (n) | 2 for 0.2 - 0.5 V range | 2.8 for 0.1 - 0.2 V range |
| | 10 for 0.5 - 1 V range | 3.7 for 0.2 - 1 V range |
| Photovoltaic mode: (At $\lambda$ = 470 and 530 nm) | | |
|     Open circuit voltage ($V_{oc}$) | 300 and 310 mV | 100 and 70 mV |
|     Short circuit current ($I_{sc}$) | 1.0 and 0.8 µA | 3.48 and 1.06 nA |
|     Fill Factor (FF) | 56 % and 64% | 32 % and 31 % |
| Photoconducting mode: (At $\lambda$ = 470 and 530 nm) | | |
| Responsivity ($R_\lambda$) | 0.069 and 0.018 A/W | 1675 and 197 A/W |
| External quantum efficiency (or) Photocurrent gain | 18.2 % and 4.2 % | $4.4 \times 10^3$ and $4.6 \times 10^2$ |